\documentclass[aps,pra,twocolumn,amsmath, pxfonts,amssymb,nofootinbib,superscriptaddress]{revtex4}

\usepackage{graphicx}

\newcommand{\bra}[1]{\langle#1|}
\newcommand{\perr}{p_{com}}
\newcommand{\ploss}{p_{loss}}
\newcommand{\pt}{p_{t}}
\newcommand{\ket}[1]{|#1\rangle}

\newcommand{\fig}[1]{Fig.~\ref{#1}}
\newcommand{\logical}{computational}

\begin{document}
\title{Thresholds for topological codes in the presence of loss}

\author{Thomas M. Stace} \email[]{stace@physics.uq.edu.au}
\affiliation{School of Mathematics and Physics, University of Queensland, Brisbane, QLD 4072, Australia}

\author{Sean D. Barrett\footnote{TMS and SDB contributed equally to this work.}} \email[]{seandbarrett@gmail.com}
\affiliation{Blackett Laboratory, Imperial College London, Prince Consort Road, London SW7 2BZ, United Kingdom}

\author{Andrew C. Doherty}
\affiliation{School of Mathematics and Physics, University of Queensland, Brisbane, QLD 4072, Australia}
\date{\today}

\frenchspacing

\begin{abstract}
Many proposals for quantum information processing are subject to detectable loss  errors. 
In this paper, we show that topological error correcting codes, which protect against \logical\ errors, are also extremely robust against losses.  
We present analytical results showing the maximum tolerable loss rate is 50\%, which is determined by the square-lattice bond percolation threshold. This saturates the bound set by the no-cloning theorem.  Our numerical results support this, and show a graceful trade-off between computational and loss errors.

\end{abstract}

\maketitle

Quantum information is delicate. Qubits can be corrupted by environmental noise, dissipation and imperfect logic gates. Quantum error correction \cite{PhysRevA.52.R2493,PhysRevLett.77.793} and fault tolerant quantum computation (FTQC) \cite{steane1999eft} address these problems, enabling quantum information to be stored and manipulated in spite of physical errors. 

 Errors may be classified as \emph \logical\  errors, in which the state of the qubits remains within the computational basis, or as  \emph{losses}, in which physical qubits (e.g. a photon) are lost from the computer.
More generally, any detectable leakage process taking a qubit out of the computational basis can be treated as a loss error.

Losses are both both \emph{detectable} and \emph{locatable}, suggesting they should be easier to rectify than computational errors.
Indeed, quantum {communication channels} can tolerate a higher rate of loss ($p_{loss}<0.5$) than depolarisation ($p_{depol}<1/3$)  \cite{PhysRevLett.78.3217}. 
 Furthermore, a quantum {computation} scheme has been developed which works as long as $\ploss < 0.5$, saturating this bound \cite{varnava:120501}. This threshold is much less restrictive than the largest estimates  for the \logical\  error threshold,  $\perr \lesssim 10^{-2}$ \cite{knill-2004,1367-2630-9-6-199}, however it is not clear how the scheme performs in the presence of both loss and \logical\ errors. Dawson et al. \cite{dawson:020501} have considered an error model which contains both loss and \logical\ errors, finding that FTQC is possible provided  $\ploss \lesssim 3\times10^{-3}$ and  $\perr \lesssim 10^{-4}$.
It is natural to ask whether alternative FTQC schemes can be found, which are tolerant to both types of error, and have less restrictive thresholds.

\begin{figure}
\begin{center}
\includegraphics[width=\columnwidth]{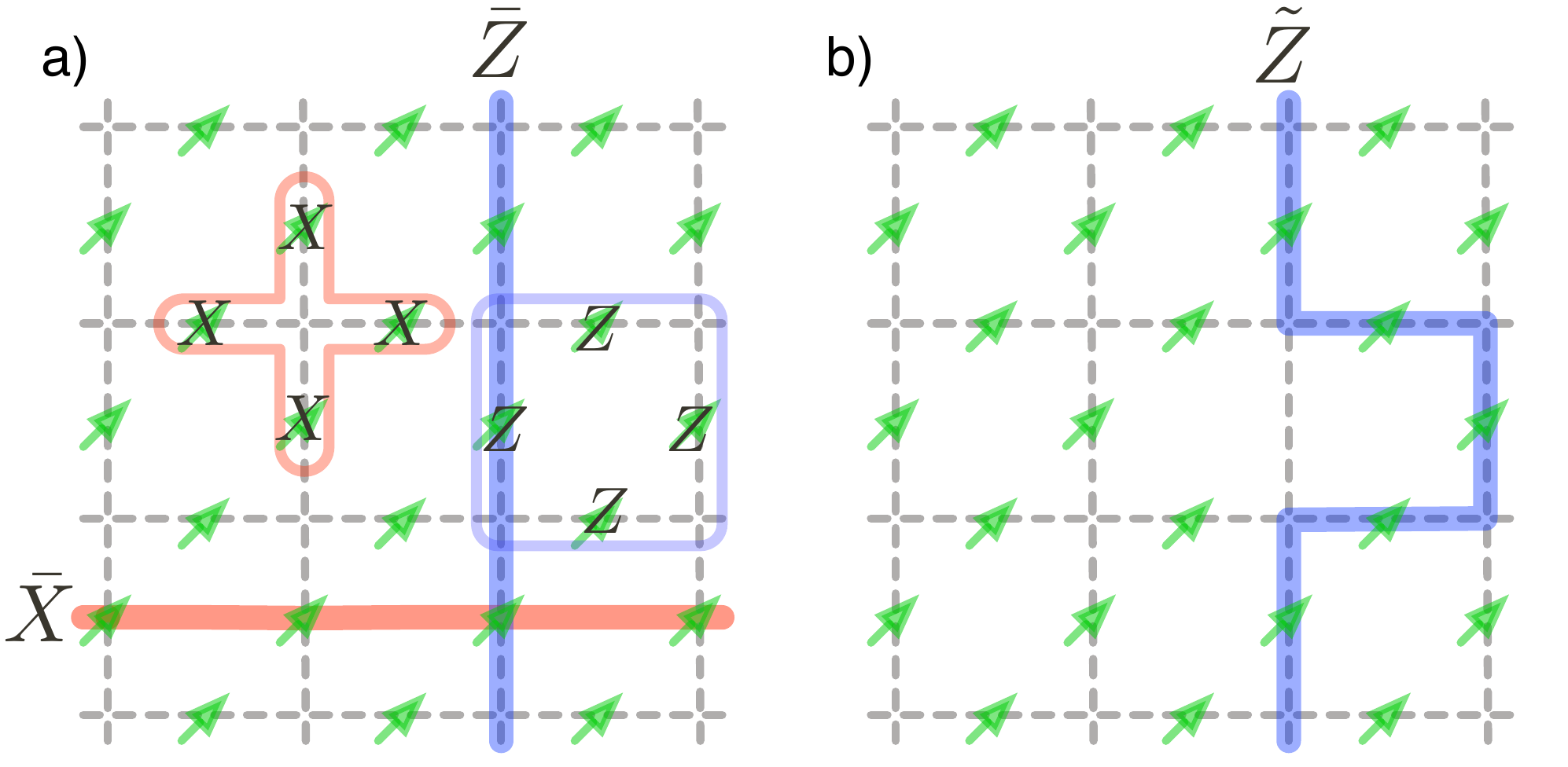}
 \caption{(a) Physical qubits (arrows) reside on the edges of a square lattice (dashed).  
 Also depicted are a plaquette operator, a star operator, the logical $\bar Z$ operator and the logical $\bar X$ operator.
  (b) In the event of a qubit loss, an equivalent logical operator $\tilde Z$ can be routed around the loss.} \label{toriccode}
\end{center}
\end{figure}

In this paper we partially address this question by considering the effect of both qubit losses and \logical\ errors on Kitaev's surface codes \cite{kitaev2003ftq}.  Surface code quantum memories are  robust against \logical\ errors, with a threshold of $\perr < 0.104$, and are an ingredient in an FTQC scheme with a high \logical\ error threshold, $\perr \lesssim 10^{-2}$ \cite{1367-2630-9-6-199}. We show that surface codes can be made robust against \emph{both} \logical\ and loss errors. Furthermore, we show that the threshold for the loss rate on the surface code is  0.5 (when $\perr = 0$) saturating known bounds. 
Here, we focus on quantum memories with idealised, perfect measurements.
We also discuss how the insights gained in this work generalise to FTQC \cite{1367-2630-9-6-199} which should therefore be robust to loss, \logical\ errors and imperfect measurement. 
 This is relevant to various implementations of cluster-state computation, including optical  \cite{Prevedel2007aa} and atomic ensembles \cite{Choi08,barrett-2008}.

For the purposes of analysis, the error model we consider is local and uncorrelated.  Each physical qubit is lost with probability $\ploss$. Losses are presumed to be detectable: a projector  onto the computational basis of a given qubit,  $\Pi_{i}=\ket{0}_i\bra{0}+\ket{1}_i\bra{1}$, is an observable indicating only whether the state of the qubit has leaked out of the computational basis.  The remaining qubits are subject to independent bit-flip ($X$) and phase  ($Z$) errors, each with probability $\perr$. Both errors are handled in an analogous way in the surface code, so here we confine our attention to $X$ errors, noting that the thresholds for $Z$ errors will be identical. Aside from these errors, we assume other quantum operations (e.g.\ syndrome measurements) can be implemented perfectly.

\begin{figure}
\begin{center}
\includegraphics[width=0.8\columnwidth]{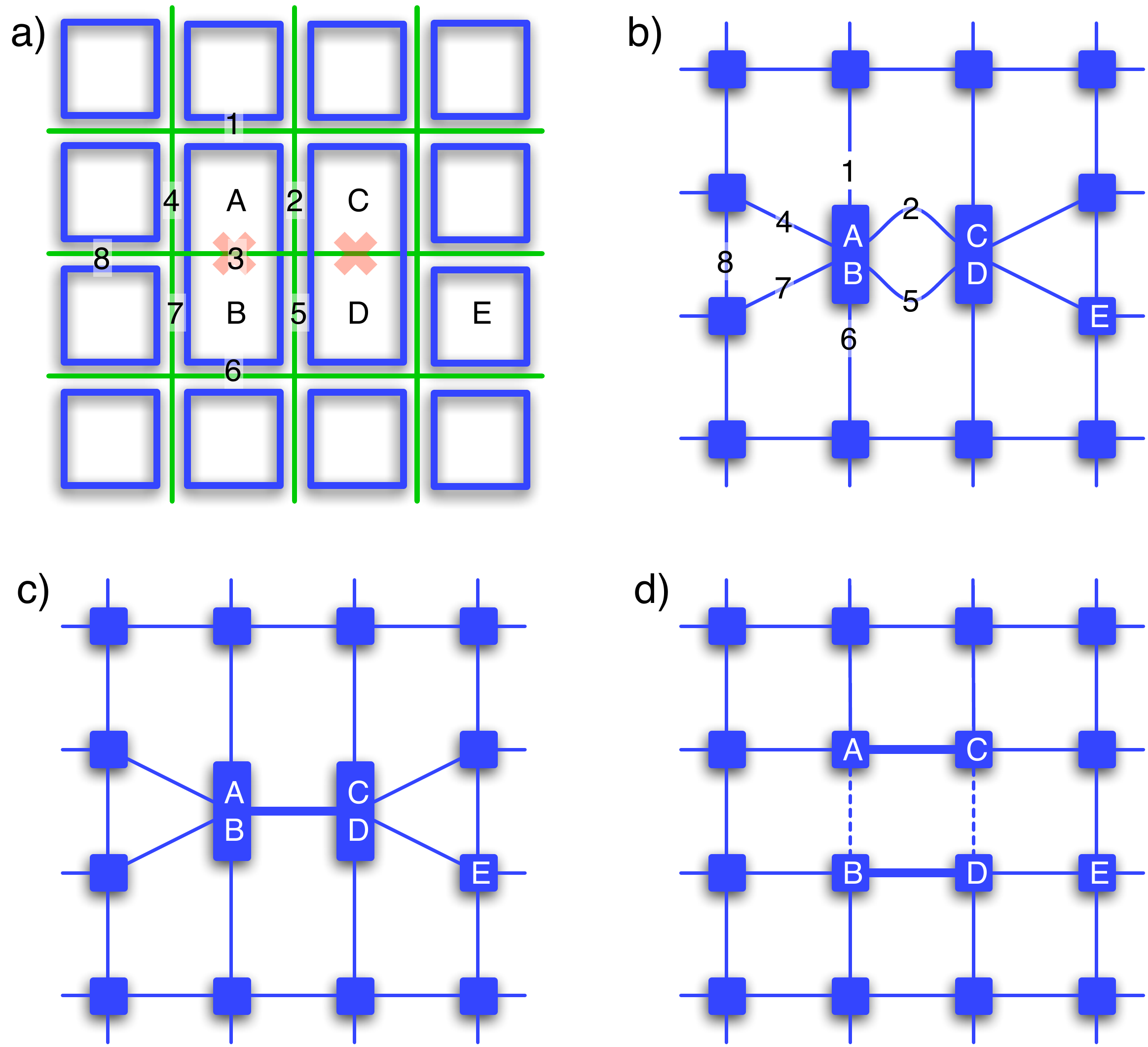}
\caption{(a) Lattice with two lost qubits (crosses). 
Representative qubits and plaquettes are labeled 1 to 8, and $\mathsf A$ to $\mathsf E$ respectively. (b) Plaquettes  sharing a lost qubit ($\{\mathsf{A}, \mathsf{B}\}$ and $\{\mathsf{C}, \mathsf{D}\}$) become superplaquettes $\mathsf{AB}$ and $\mathsf{CD}$, and may be multiply connected (i.e.\ share more than one qubit). (c) Degraded lattice showing superedges (thick lines).   (d) Restored lattice with zero weight edges (dotted) and irregular weights (thick).} \label{multiedges}
\end{center}
\end{figure}

Kitaev's surface codes are defined by a set of stabilisers   acting on a set of physical qubits that reside on the edges of a square lattice \cite{PhysRevA.57.127}.  The stabiliser group is generated by plaquette operators, which are products of $Z$ operators acting on qubits within a plaquette, $P_p =  \bigotimes_{i \in p} Z_{i}$, and star operators, which are products of $X$ operators acting on qubits within a star, $S_s = \bigotimes_{j \in s} X_{j}$, as depicted in \fig{toriccode} \cite{kitaev2003ftq}.  The stabilisers  commute, and the code space $\{|C\rangle\}$ is a simultaneous $+1$ eigenstate of all stabilisers. If the $L\times L$ lattice has periodic boundary conditions, there are $2L^{2}$ physical qubits and $2(L^2-1)$ independent stabilisers.  The two remaining degrees of freedom are capable of encoding two logical qubits, $\bar q_i$  ($i \in \{1,2\}$).  A logical $\bar{Z}_i$ ($\bar{X}_i$) operator corresponds to a product of $Z$ ($X$) operators along a homologically nontrivial cycle (i.e.\ spanning the lattice), shown in \fig{toriccode}.  $\bar{Z}_i$  and $\bar{X}_i$  commute with the stabilisers, but are not contained within the stabiliser group.

The logical operators are not unique; e.g.\ multiplying  $\bar Z_i$ by a plaquette stabiliser yields a new operator $\tilde{Z}_i = P_p \bar{Z}_i$ that acts  identically  on the code space, \mbox{$\tilde{Z}_i |C\rangle = P_p \bar{Z}_i |C\rangle =   \bar{Z}_i  P_p|C\rangle =   \bar{Z}_i |C\rangle$}. Thus there are many homologically equivalent cycles with which to measure each logical qubit operator, as shown in \fig{toriccode}(b). This redundancy allows us to obtain the loss threshold for the case $\perr=0$: if only a few physical qubits are lost, it is likely that each logical operator can be reliably measured by finding a homologically nontrival cycle that avoids all lost qubits. 

If  $\ploss$ is too high, there is likely to be a \emph{percolated} region of losses spanning the entire lattice, in which case there are no homologically nontrivial cycles with which to measure the logical operators. As $L\rightarrow\infty$, 
there is a sharp boundary between recoverable and non-recoverable errors 
corresponding to the \emph{bond percolation threshold} on the square lattice \cite{Stauffer1985aa}: for $\ploss<0.5$ loss recovery almost surely succeeds, whereas for  $\ploss>0.5$ loss recovery almost surely fails. Notably, this threshold saturates the bound on $\ploss$  imposed by the no-cloning theorem \cite{PhysRevLett.78.3217}.

The case  $\ploss=0$  and $\perr>0$ has been well studied \cite{kitaev2003ftq,dennis:4452,wang2003cht}. Briefly, physical bit-flip errors lead to logical bit-flip ($\bar{X}_i$) errors but not logical phase  errors, and vice-versa. An \emph{error chain}, $E$, is a set of lattice edges (i.e.\ physical qubits) where a bit-flip error has occurred. 
The plaquette operator eigenvalues change to $-1$ only at the \emph{boundary}, $\partial{E}$, of the chain.   
Measuring the plaquette operators therefore yields information about the endpoints of connected sub-chains of $E$. If $E$ crosses $\bar{Z}_i$ an odd number of times, then the logical qubit suffers a $\bar{X}_i$ error. These errors may be corrected if the endpoints, $\partial E$, can be matched by a correction chain, $E'$, such that the closed chain $C=E+E'$ crosses  $\bar{Z}_i$ an even number of times, i.e.\ $C$ is homologially trivial.  The error rate below which the correction chain $E'$ may be successfully constructed is closely related to the phase boundary of the random-bond Ising model (RBIM) \cite{wang2003cht,PhysRevE.64.046108}. If $\perr<p_{c0}=0.104$ \cite{footnote1},
 then in the limit $L\rightarrow\infty$, the most probable chain, $C_{max}=E+E'_{max}$, is almost surely homologically trivial and recovery succeeds.  If $\perr>p_{c0}$, then in the limit $L\rightarrow\infty$, the chain is homologically trival only $25\%$ of the time, and recovery fails.  
 
We can think of the above results as endpoints of a `boundary of correctability': $(\ploss,\perr)=(0.5,0)$ and $(0,0.104)$, respectively.   In what follows, we demonstrate that toric codes (and planar codes, by extension) are robust against both loss and \logical\ errors, with a graceful tradeoff between the two.  We first describe how losses can be corrected by forming new stabiliser generators, which are aggregations of plaquettes or stars, called superplaquettes and superstars, respectively.  The superstar and superplaquette eigenvalues then reveal the error syndromes, and a  perfect matching algorithm  is used to find an error correction chain $E'$.
We illustrate the efficacy of the single round error correction protocol by calculating numerically the boundary of correctability in the $(\ploss, \perr)$ parameter space.

Consider the lattice shown in \fig{multiedges}(a) which is damaged by the loss of two physical qubits, marked by the crosses.  The loss of qubit 3 affects two plaquette stabilisers: $P_\mathsf{A}=Z_1Z_2Z_3Z_4$ and $P_ \mathsf B=Z_3Z_5Z_6Z_7$, rendering them  unmeasurable.  
However, the \emph{superplaquette} $P_ \mathsf{AB}=P_ \mathsf AP_ \mathsf B=Z_1Z_2Z_4Z_5Z_6Z_7$ is independent of the qubit at site 3, so  stablises  the remaining qubits.   Without  errors, $P_ \mathsf{AB}$ has an eigenvlaue of $+1$. An error chain ending within the superplaquette $\mathsf {AB}$ changes the eigenvalue of $P_ \mathsf {AB}$ to $-1$.  
It follows that  the syndrome associated with a superplaquette is determined by the parity of the number of error chains that cross its boundary.  The fact that superplaquette operators yield syndrome information with which to construct an error correction chain, $E'$, is the basis for our loss-tolerant error-correction scheme.  

 In general, given any set of lost qubits, we can form a complete set of stabilisers on the damaged lattice in the following way: for each lost qubit $q$, which participates in neighbouring (super)plaquettes $P_q$ and $P'_q$, we form the superplaquette operator $P_qP'_q$, which is  independent of $Z_q$.  In the same way, we form superstar operators from products of star operators.  As discussed earlier, we can also form new logical $\tilde X_i$ and $\tilde Z_i$ operators by deforming the original logical operators to conform to the boundaries of newly formed superplaquettes.

We note that in \fig{multiedges}(a), there is a damaged plaquette operator $\bar Z_\mathcal{J}=Z_3P_\mathsf{A}=Z_1Z_2Z_4$ (or, equivalently $Z_3P_ \mathsf B=Z_5Z_6Z_7$) associated with the lost qubit 3,  which commutes with all the newly formed stabiliser generators on the damaged lattice, but whose eigenvalue, $\pm1$, is indeterminate.  Likewise, the damaged star operator $\bar X_\mathcal{J}=X_4X_7X_8$ also has indeterminate eigenvalue and commutes with the new stabilisers on the damaged lattice.  Having indeterminate eigenvalues, $\bar Z_\mathcal{J}$ and $\bar X_\mathcal{J}$, which mutually anticommute, define a two-dimensional degree of freedom in an uncertain state.  They therefore describe a completely mixed \emph{junk} qubit, $ \mathcal{J}$,  which is a consequence of the entanglement between the lost qubit and the remaining qubits \cite{PhysRevA.71.022315}.  Since $\bar Z_{ \mathcal{J}}$ and $\bar X_{ \mathcal{J}}$ each commute with the new stabilisers, and with the deformed logical operators, the junk qubit is in a product state with the logical qubits: $\ket{\psi}\bra{\psi}_{\bar q_1}\otimes\ket{\phi}\bra{\phi}_{\bar q_2}\otimes\mathbb{I}_{ \mathcal{J}}/2$, and so the loss does not affect the logical qubits.

When analysing the pattern of syndromes on the plaquettes and superplaquettes, we construct a new graph, depicted in \fig{multiedges}(b), in which a (super)plaquette is represented by a node, and (super)plaquettes share a bond on the new graph wherever they share a physical qubit in common.  Thus $P_ \mathsf{AB}$ and $P_ \mathsf{CD}$ share the qubits 2 and 5, and this is represented as two edges between the superplaquette nodes labelled $\mathsf {AB}$ and $\mathsf {CD}$.

The error correction  syndrome, $\partial E$, arising from an error chain on the graph in \fig{multiedges}(b) is determined by the (super)plaquettes that have an eigenvalue of $-1$.  To correct the errors, we follow the procedure described by Harrington et al \cite{dennis:4452, wang2003cht} to find the most likely error chain giving rise to $\partial E$. 
The probability of a given error chain is modified by the presence of losses.  With no loss, the probability of an error on a qubit, $\ell=\{\mathsf {P,P'}\}$,  between two neighbouring plaquettes $\mathsf P$ and $\mathsf P'$,  is uniform, $p_\ell=\perr$.    With loss, superplaquettes may share multiple physical qubits in common, as shown in \fig{multiedges}(b), where  superplaquettes $\mathsf {AB}$ and $\mathsf {CD}$ have qubits 2 and 5 in common. A non-trivial syndrome arises only if either qubit 2 or qubit 5 suffers an error, but not both.
By extension, for a pair of neighbouring superplaquettes, $\ell=\{\mathsf P, \mathsf P'\}$, sharing $n_{\ell}$ physical qubits, a non-trivial syndrome arises only if there are an odd number of errors on the $n_\ell$ qubits, which happens with probability 
\begin{equation}
p_\ell=\sum_{m\textrm{ odd}}^{n_\ell} {n_\ell \choose m}\perr^m(1-\perr)^{n_\ell-m}=\frac{1-(1-2\perr)^{n_\ell}}{2}\nonumber
\end{equation}

We therefore \emph{degrade} the graph shown in \fig{multiedges}(b), replacing multi-edges (i.e.\ several shared physical qubits) whose error probabilities are uniform,  with single \emph{superedges} whose error rates  depend on the number of physical qubits shared between neighbouring superplaquettes.  This degraded lattice is shown in \fig{multiedges}(c), in which there are no multi-edges, but the error probabilities are no longer constant.  

On this degraded lattice, we may now assign a probability for any hypothetical chain $E'=E+C$, where $C$ is a closed chain.  This probability, which is conditioned on the measured syndrome, $\partial E$ is  \cite{dennis:4452,wang2003cht}
\begin{equation}
P(E'|\partial E)=\mathcal{N}\prod_{\forall\ell}e^{J_\ell u_\ell^{E'}},\textrm{ where }u_\ell^{\mathcal{C}}=\left\{\begin{array}{cc}-1 & \textrm{if }\ell\in \mathcal{C}\\+1 & \textrm{if }\ell\notin \mathcal{C}\end{array}\right.\nonumber
\end{equation}
for a chain $\mathcal{C}$,   $\mathcal{N}=\prod_{\forall\ell}\sqrt{p_\ell(1-p_\ell)}$ is a normalisation constant and $e^{2J_\ell}=1/p_\ell-1$. 

The chain $E'$ that maximises $P(E'|\partial E)$ also minimises
$
\sum_{\ell\in E'} J_\ell.
$
This minimisation may be accomplished using Edmonds' minimum-weight, perfect-matching algorithm \cite{WilliamCook01011999}.  
For  $\ploss=0$, this simply minimises the total metropolis length of the matching path, and is the same procedure implemented in previous studies \cite{dennis:4452, wang2003cht}.  For $\ploss>0$, the edge weights are not  uniform, since $p_\ell$ depends on the number of physical qubits, $n_\ell$, shared between adjacent superplaquettes.

For the purposes of simulation,  it is easier to determine homology classes on a square lattice, rather than the degraded lattice, exemplified in  \fig{multiedges}(c).  We therefore restore the square lattice by dividing superplaquettes into their constituent plaquettes in the following way:  (1) an edge between two plaquettes  within a single superplaquette is assigned a weight of zero, (2) an edge between plaquettes in two neighbouring superplaquettes is given the weight of the superedge in the degraded lattice, as illustrated in \fig{multiedges}(d).  These transformations do not change the weighted distance between any pair of syndromes, and so a minimum-weight perfect matching on the restored lattice is also a minimum-weight perfect matching on the degraded lattice.  Determining the homology class is then accomplished by counting crossings of  vertical and horizontal test lines in the dual lattice.

In order to test the efficacy of our loss-tolerant error correction scheme, we generate random losses on a periodic lattice with rate $\ploss$.   On the remaining qubits we generate a set of errors, $E$, with rate $\perr$. 
Applying Edmonds'  algorithm to $\partial E$ on the weighted lattice  yields the maximum-likelihood error correction chain, $E'$.  The homology class of the chain $E+E'$ then determines whether error correction was successful.  

\begin{figure}
\begin{center}
\includegraphics[width=\columnwidth]{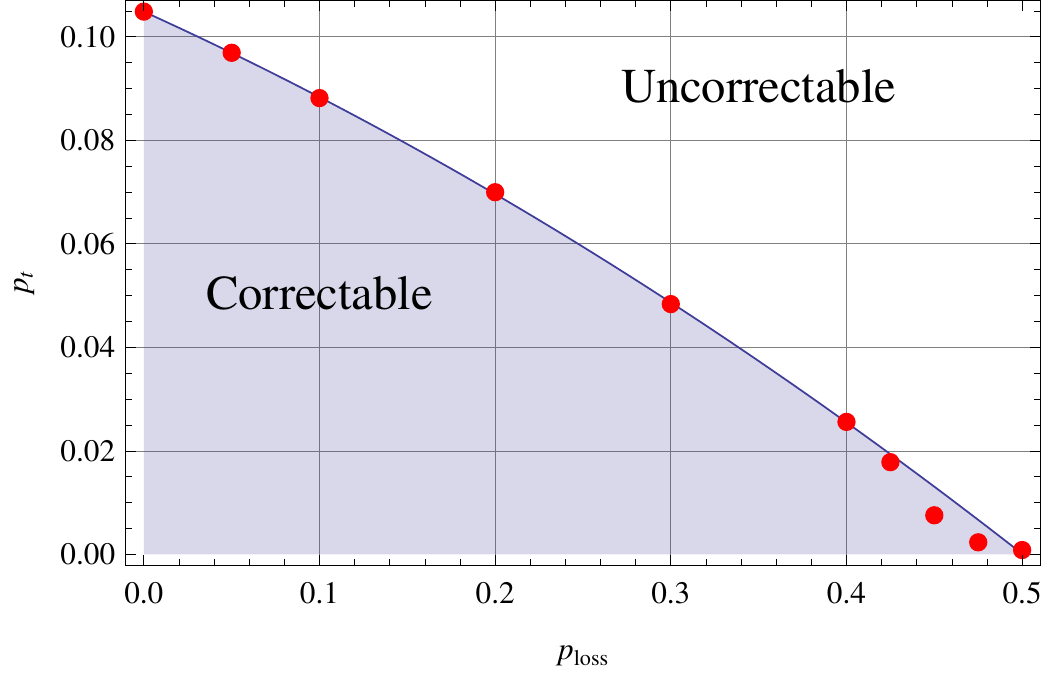}
\caption{ Correctability phase diagram.  The shaded region is correctable in the limit $L\rightarrow\infty$. The threshold, $p_t$, is calculated by fitting the universal scaling law  $p_{fail}=f[(\perr-\pt)L^{1/\nu_0}]$.
The curve is a quadratic fit to the points for which $\ploss\leq0.4$ (where universal scaling is unaffected by the finite lattice size). It extrapolates through $(\ploss,\pt)=(0.5,0)$.} \label{PhaseDiagram}
\end{center}
\end{figure}

For each value of $\ploss$ we simulate the protocol for different values of $\perr$ on lattice sizes, $L=16, 24$ and 32.  For given values of $\perr$ and $L$, the failure rate, $p_{fail}$, is calculated by averaging over $10^4$ trials. 
Following \cite{wang2003cht}, we seek a threshold, $\pt$ (depending on $\ploss$), such that $d p_{fail}/dL<0$ when $\perr<\pt$, and $d p_{fail}/dL>0$ when  $\perr>\pt$.  
That is, for each value of $\ploss$, we fit the simulated failure rate to a universal scaling law $p_{fail}=f[x]$ ($\approx a+b x$, for small $x$), where \mbox{$x=(\perr-\pt)L^{1/\nu_0}$}, with fitting parameters $p_t, \nu_0, a$ and $b$. 

\fig{PhaseDiagram} is the central result in this paper, and shows $\pt$ as a function of $\ploss$.   For the points $\ploss\leq0.4$, the  universal scaling law provides a good fit to the simulated results, so $\pt$ is well defined, and $\nu_0\approx1.5$ consistent with  the scaling exponent of the RBIM universality class \cite{wang2003cht}.    A quadratic fit through these points yields a curve that extrapolates through  $(0.5,0)$.  This curve represents the boundary of correctability: if $(\ploss,\perr)$ is in the shaded region then the failure rate decreases to zero as $L$ increases.  Importantly, this boundary  passes through the known bounds at $\ploss=0$ and $0.5$, demonstrating that the protocol is very robust against loss.

For $\ploss\geq0.425$, the universal scaling assumption breaks down (best-fits for $p_t$ are still shown), and the points in \fig{PhaseDiagram} lie below the quadratic extrapolation (but still attaining the point $(0.5,0)$).  This is attributed to the fact that for $\ploss\geq0.425$, the largest superplaquette on an $L\leq32\times32$ lattice  occupies approximately half of the lattice sites \cite{PhysRevE.62.1660}, so finite size effects dominate. 

The protocol described in this paper for dealing with losses in a surface code relies on several important properties of the stabilisers.  Firstly, if a physical qubit $q$ in the logical qubit operator chain $\bar Z_i$ is lost, then there is a plaquette $P_q$, such that $\tilde Z_i =\bar Z_i P_q$ is independent of $q$ (likewise $\bar X_i\rightarrow\tilde X_i$).  Thus, logical operator chains can be rerouted around the lost site. Secondly, there is another plaquette $P'_q$  such that the superplaquette $P'_qP_q$ is independent of $q$. Thus, superplaquettes may be constructed to locate the endpoints of error chains  (likewise for superstars). Thirdly, newly formed junk qubits are uncorrelated with the logical qubits.

These properties are satisfied by a number of related protocols, including the fault-tolerant planar code \cite{dennis:4452, wang2003cht}, long-range entanglement creation in a noisy system \cite{raussendorf:062313}, and in Raussendorf's topological, fault-tolerant, cluster-state quantum computation scheme \cite{1367-2630-9-6-199}, 
 which borrows a number of ideas from anyonic quantum computation.  In these protocols, the syndrome operators exhibit simple cubic symmetry, whose bond  percolation threshold is $p_{sc}\approx0.249$ \cite{Stauffer1985aa}.  We therefore expect that the region of correctability for these schemes includes the points  $(\ploss,\perr)=(0,0.029)$ and $(0.249,0)$.
 
We have demonstrated that surface codes are robust to noise arising from both errors and losses.  The correctable phase in $(\ploss,\perr)$ space includes the known results, and we have shown that for a model of uncorrelated noise, there is a significant fraction of parameter space in which the surface code is robust to both loss and computational errors.  The approach described here is applicable to other systems, including full, fault-tolerant quantum computation, as well as correlated noise models.  

TMS and ACD thank the 
ARC for funding.  SDB thanks the EPSRC for funding.  We thank J.\ Harrington, V.\ Kolmogorov and W.\ Cook for helpful comments.


\begin{thebibliography}{22}
\expandafter\ifx\csname natexlab\endcsname\relax\def\natexlab#1{#1}\fi
\expandafter\ifx\csname bibnamefont\endcsname\relax
  \def\bibnamefont#1{#1}\fi
\expandafter\ifx\csname bibfnamefont\endcsname\relax
  \def\bibfnamefont#1{#1}\fi
\expandafter\ifx\csname citenamefont\endcsname\relax
  \def\citenamefont#1{#1}\fi
\expandafter\ifx\csname url\endcsname\relax
  \def\url#1{\texttt{#1}}\fi
\expandafter\ifx\csname urlprefix\endcsname\relax\def\urlprefix{URL }\fi
\providecommand{\bibinfo}[2]{#2}
\providecommand{\eprint}[2][]{\url{#2}}

\bibitem[{\citenamefont{Shor}(1995)}]{PhysRevA.52.R2493}
\bibinfo{author}{\bibfnamefont{P.~W.} \bibnamefont{Shor}},
  \bibinfo{journal}{Phys. Rev. A} \textbf{\bibinfo{volume}{52}},
  \bibinfo{pages}{R2493} (\bibinfo{year}{1995}).

\bibitem[{\citenamefont{Steane}(1996)}]{PhysRevLett.77.793}
\bibinfo{author}{\bibfnamefont{A.~M.} \bibnamefont{Steane}},
  \bibinfo{journal}{Phys. Rev. Lett.} \textbf{\bibinfo{volume}{77}},
  \bibinfo{pages}{793} (\bibinfo{year}{1996}).

\bibitem[{\citenamefont{Steane}(1999)}]{steane1999eft}
\bibinfo{author}{\bibfnamefont{A.}~\bibnamefont{Steane}},
  \bibinfo{journal}{Nature} \textbf{\bibinfo{volume}{399}},
  \bibinfo{pages}{124} (\bibinfo{year}{1999}).

\bibitem[{\citenamefont{Bennett et~al.}(1997)\citenamefont{Bennett, DiVincenzo,
  and Smolin}}]{PhysRevLett.78.3217}
\bibinfo{author}{\bibfnamefont{C.~H.} \bibnamefont{Bennett}},
  \bibinfo{author}{\bibfnamefont{D.~P.} \bibnamefont{DiVincenzo}},
  \bibnamefont{and} \bibinfo{author}{\bibfnamefont{J.~A.}
  \bibnamefont{Smolin}}, \bibinfo{journal}{Phys. Rev. Lett.}
  \textbf{\bibinfo{volume}{78}}, \bibinfo{pages}{3217} (\bibinfo{year}{1997}).

\bibitem[{\citenamefont{Varnava et~al.}(2006)\citenamefont{Varnava, Browne, and
  Rudolph}}]{varnava:120501}
\bibinfo{author}{\bibfnamefont{M.}~\bibnamefont{Varnava}},
  \bibinfo{author}{\bibfnamefont{D.~E.} \bibnamefont{Browne}},
  \bibnamefont{and} \bibinfo{author}{\bibfnamefont{T.}~\bibnamefont{Rudolph}},
  \bibinfo{journal}{Phys.\ Rev.\ Lett.} \textbf{\bibinfo{volume}{97}},
  \bibinfo{eid}{120501} (\bibinfo{year}{2006}).

\bibitem[{\citenamefont{Knill}(2004)}]{knill-2004}
\bibinfo{author}{\bibfnamefont{E.}~\bibnamefont{Knill}},
  \bibinfo{journal}{quant-ph/0404104}  (\bibinfo{year}{2004}).

\bibitem[{\citenamefont{Raussendorf et~al.}(2007)\citenamefont{Raussendorf,
  Harrington, and Goyal}}]{1367-2630-9-6-199}
\bibinfo{author}{\bibfnamefont{R.}~\bibnamefont{Raussendorf}},
  \bibinfo{author}{\bibfnamefont{J.}~\bibnamefont{Harrington}},
  \bibnamefont{and} \bibinfo{author}{\bibfnamefont{K.}~\bibnamefont{Goyal}},
  \bibinfo{journal}{New J.\ Phys.} \textbf{\bibinfo{volume}{9}},
  \bibinfo{pages}{199} (\bibinfo{year}{2007}).

\bibitem[{\citenamefont{Dawson et~al.}(2006)\citenamefont{Dawson, Haselgrove,
  and Nielsen}}]{dawson:020501}
\bibinfo{author}{\bibfnamefont{C.~M.} \bibnamefont{Dawson}},
  \bibinfo{author}{\bibfnamefont{H.~L.} \bibnamefont{Haselgrove}},
  \bibnamefont{and} \bibinfo{author}{\bibfnamefont{M.~A.}
  \bibnamefont{Nielsen}}, \bibinfo{journal}{Phys.\ Rev.\ Lett.}
  \textbf{\bibinfo{volume}{96}}, \bibinfo{eid}{020501} (\bibinfo{year}{2006}).

\bibitem[{\citenamefont{Kitaev}(2003)}]{kitaev2003ftq}
\bibinfo{author}{\bibfnamefont{A.}~\bibnamefont{Kitaev}},
  \bibinfo{journal}{Annals of Physics} \textbf{\bibinfo{volume}{303}},
  \bibinfo{pages}{2} (\bibinfo{year}{2003}).

\bibitem[{\citenamefont{Prevedel et~al.}(2007)\citenamefont{Prevedel, Walther,
  Tiefenbacher, Bohi, Kaltenbaek, Jennewein, and Zeilinger}}]{Prevedel2007aa}
\bibinfo{author}{\bibfnamefont{R.}~\bibnamefont{Prevedel}},
  \bibinfo{author}{\bibfnamefont{P.}~\bibnamefont{Walther}},
  \bibinfo{author}{\bibfnamefont{F.}~\bibnamefont{Tiefenbacher}},
  \bibinfo{author}{\bibfnamefont{P.}~\bibnamefont{Bohi}},
  \bibinfo{author}{\bibfnamefont{R.}~\bibnamefont{Kaltenbaek}},
  \bibinfo{author}{\bibfnamefont{T.}~\bibnamefont{Jennewein}},
  \bibnamefont{and}
  \bibinfo{author}{\bibfnamefont{A.}~\bibnamefont{Zeilinger}},
  \bibinfo{journal}{Nature} \textbf{\bibinfo{volume}{445}}, \bibinfo{pages}{65}
  (\bibinfo{year}{2007}).

\bibitem[{\citenamefont{Choi et~al.}(2008)\citenamefont{Choi, Deng, Laurat, and
  Kimble}}]{Choi08}
\bibinfo{author}{\bibfnamefont{K.~S.} \bibnamefont{Choi}},
  \bibinfo{author}{\bibfnamefont{H.}~\bibnamefont{Deng}},
  \bibinfo{author}{\bibfnamefont{J.}~\bibnamefont{Laurat}}, \bibnamefont{and}
  \bibinfo{author}{\bibfnamefont{H.~J.} \bibnamefont{Kimble}},
  \bibinfo{journal}{Nature} \textbf{\bibinfo{volume}{452}}, \bibinfo{pages}{67}
  (\bibinfo{year}{2008}).

\bibitem[{\citenamefont{Barrett et~al.}(2008)\citenamefont{Barrett, Rohde, and
  Stace}}]{barrett-2008}
\bibinfo{author}{\bibfnamefont{S.~D.} \bibnamefont{Barrett}},
  \bibinfo{author}{\bibfnamefont{P.~P.} \bibnamefont{Rohde}}, \bibnamefont{and}
  \bibinfo{author}{\bibfnamefont{T.~M.} \bibnamefont{Stace}},
  \bibinfo{journal}{arXiv:0804.0962}  (\bibinfo{year}{2008}).

\bibitem[{\citenamefont{Gottesman}(1998)}]{PhysRevA.57.127}
\bibinfo{author}{\bibfnamefont{D.}~\bibnamefont{Gottesman}},
  \bibinfo{journal}{Phys. Rev. A} \textbf{\bibinfo{volume}{57}},
  \bibinfo{pages}{127} (\bibinfo{year}{1998}).

\bibitem[{\citenamefont{Stauffer}(1985)}]{Stauffer1985aa}
\bibinfo{author}{\bibfnamefont{D.}~\bibnamefont{Stauffer}},
  \emph{\bibinfo{title}{Introduction to Percolation Theory}}
  (\bibinfo{publisher}{Taylor and Francis}, \bibinfo{year}{1985}).

\bibitem[{\citenamefont{Wang et~al.}(2003)\citenamefont{Wang, Harrington, and
  Preskill}}]{wang2003cht}
\bibinfo{author}{\bibfnamefont{C.}~\bibnamefont{Wang}},
  \bibinfo{author}{\bibfnamefont{J.}~\bibnamefont{Harrington}},
  \bibnamefont{and} \bibinfo{author}{\bibfnamefont{J.}~\bibnamefont{Preskill}},
  \bibinfo{journal}{Annals of Physics} \textbf{\bibinfo{volume}{303}},
  \bibinfo{pages}{31} (\bibinfo{year}{2003}).

\bibitem[{\citenamefont{Dennis et~al.}(2002)\citenamefont{Dennis, Kitaev,
  Landahl, and Preskill}}]{dennis:4452}
\bibinfo{author}{\bibfnamefont{E.}~\bibnamefont{Dennis}},
  \bibinfo{author}{\bibfnamefont{A.}~\bibnamefont{Kitaev}},
  \bibinfo{author}{\bibfnamefont{A.}~\bibnamefont{Landahl}}, \bibnamefont{and}
  \bibinfo{author}{\bibfnamefont{J.}~\bibnamefont{Preskill}},
  \bibinfo{journal}{J. Math. Phys.} \textbf{\bibinfo{volume}{43}},
  \bibinfo{pages}{4452} (\bibinfo{year}{2002}).

\bibitem[{\citenamefont{Nobre}(2001)}]{PhysRevE.64.046108}
\bibinfo{author}{\bibfnamefont{F.~D.} \bibnamefont{Nobre}},
  \bibinfo{journal}{Phys. Rev. E} \textbf{\bibinfo{volume}{64}},
  \bibinfo{pages}{046108} (\bibinfo{year}{2001}).

\bibitem[{foo()}]{footnote1}
\bibinfo{note}{{Some studies give $p_{c0}=0.1031$ \cite{wang2003cht} whilst
  others give $p_{c0}=0.1049$ \cite{PhysRevE.64.046108}.}}

\bibitem[{\citenamefont{Hamma et~al.}(2005)\citenamefont{Hamma, Ionicioiu, and
  Zanardi}}]{PhysRevA.71.022315}
\bibinfo{author}{\bibfnamefont{A.}~\bibnamefont{Hamma}},
  \bibinfo{author}{\bibfnamefont{R.}~\bibnamefont{Ionicioiu}},
  \bibnamefont{and} \bibinfo{author}{\bibfnamefont{P.}~\bibnamefont{Zanardi}},
  \bibinfo{journal}{Phys. Rev. A} \textbf{\bibinfo{volume}{71}},
  \bibinfo{pages}{022315} (\bibinfo{year}{2005}).

\bibitem[{\citenamefont{Cook and Rohe}(1999)}]{WilliamCook01011999}
\bibinfo{author}{\bibfnamefont{W.}~\bibnamefont{Cook}} \bibnamefont{and}
  \bibinfo{author}{\bibfnamefont{A.}~\bibnamefont{Rohe}},
  \bibinfo{journal}{Informs J. Comp.} \textbf{\bibinfo{volume}{11}},
  \bibinfo{pages}{138} (\bibinfo{year}{1999}).

\bibitem[{\citenamefont{Bazant}(2000)}]{PhysRevE.62.1660}
\bibinfo{author}{\bibfnamefont{M.~Z.} \bibnamefont{Bazant}},
  \bibinfo{journal}{Phys. Rev. E} \textbf{\bibinfo{volume}{62}},
  \bibinfo{pages}{1660} (\bibinfo{year}{2000}).

\bibitem[{\citenamefont{Raussendorf et~al.}(2005)\citenamefont{Raussendorf,
  Bravyi, and Harrington}}]{raussendorf:062313}
\bibinfo{author}{\bibfnamefont{R.}~\bibnamefont{Raussendorf}},
  \bibinfo{author}{\bibfnamefont{S.}~\bibnamefont{Bravyi}}, \bibnamefont{and}
  \bibinfo{author}{\bibfnamefont{J.}~\bibnamefont{Harrington}},
  \bibinfo{journal}{Phys.\ Rev.\ A} \textbf{\bibinfo{volume}{71}},
  \bibinfo{eid}{062313} (\bibinfo{year}{2005}).

\end{thebibliography}

\end{document}